\documentclass[12pt]{article}
\usepackage[cp1251]{inputenc}
\usepackage{amsmath}
\usepackage{amssymb}
\usepackage[english]{babel}
\usepackage{epsf}
\usepackage{pazha}
\tightenlines

\sloppypar

\begin{document}
\baselineskip 21pt

\begin{flushleft}
\textit{Astronomy Letters, 2016, Vol. 42, No. 11, pp. 745--751. \copyright~Pleiades Publishing, Inc., 2016.}
\end{flushleft}

\title{\bf BREAKDOWN OF THE GOLDREICH-JULIAN RELATION\\IN A NEUTRON STAR}

\author{\bf D. N. Sob'yanin\affilmark{*}}

\affil{
{\it Tamm Division of Theoretical Physics,\\Lebedev Physical Institute, Russian Academy of Sciences,\\Leninskii pr. 53, Moscow, 119991 Russia}}

\begin{center}
Received February 16, 2016
\end{center}

\sloppypar
\vspace{2mm}
\noindent
The electromagnetic field in a magnetized neutron star and the underlying volume charges and currents are found. A general case of a rigidly rotating neutron star with infinite conductivity, arbitrary distribution of the internal magnetic field, arbitrarily changing angular velocity, and arbitrary surface velocity less than the velocity of light is considered. Quaternions are used to describe rotation and determine the magnetic field. It is shown that the charge density is not equal to and can exceed significantly the common Goldreich-Julian density. Moreover, corrections to the magnetic field due to stellar rotation are zero. For a rotating neutron star, twisting magnetic field lines causes charge accumulation and current flows. This fact shows a possible link between changing internal magnetic field topology and observed activity of neutron stars.


\vfill
\noindent\rule{8cm}{1pt}\\
{$^*$ E-mail $<$sobyanin@lpi.ru$>$}

\clearpage

\section*{INTRODUCTION}
\noindent
It is well known that the classical observational manifestation of the neutron star is the radio pulsar (Beskin 1999). Other interesting manifestations have also been actively studied in recent years, e.g., magnetars (Mereghetti 2008), gamma-ray pulsars (Caraveo 2014), rotating radio transients (RRATs) (McLaughlin et~al. 2006), extreme nullers (Wang et~al. 2007), and even hybrids of the above objects (Burke-Spolaor and Bailes 2010). Such diverse manifestations of neutron stars are related to the changing activity of the neutron star exterior, a magnetosphere, and show that the magnetosphere can be not only purely vacuum or completely filled with a plasma but also essentially nonstationary, when the state of some its regions constantly changes from vacuum to plasma-filled and vice versa (Istomin and Sob'yanin 2011a, 2011b, 2011c).

Meanwhile, properties of the neutron star interior can also be reflected in various observable effects, with precession (Link 2007), glitches (Espinoza et~al. 2011), and bursts (Deibel et~al. 2014) as some examples. The latter effects can be related to deformation of the neutron star (Duncan 1998; Makishima et~al. 2014; Haskell and Melatos 2015). The internal magnetic field plays significant role as magnetic stresses contribute to deformation and can indirectly, via mechanic effects, change observational properties of neutron stars (Cutler 2002; Lander et~al. 2015; Garc\'{\i}a and Ranea-Sandoval 2015). Can there be other, not only mechanical, manifestations of the internal magnetic field? A possibility is generation of currents by a changing field, which can reveal itself, e.g., via crust heating and concomitant X-ray and radio emission. The currents are caused by charge redistribution and can be high if the charge density may change and reach considerable values. However, it is widely assumed that the charge density is bounded by a Goldreich-Julian density and, moreover, just equal to it (Goldreich and Julian 1969). The purpose of this paper is to show that in actual fact this is not the case.

For this purpose, I calculate the internal electromagnetic fields, charges, and currents for a general case of a rigidly rotating perfectly conducting magnetized neutron star. I consider the relation between charges and currents and show that the internal charge density is not and can exceed significantly the Goldreich-Julian density. Charge redistribution and accumulation can be realized via twisting magnetic field lines. As a consequence, internal magnetic field rearrangement can potentially be reflected in observational properties of neutron stars even without contribution from additional mechanic effects.

\section*{INTERNAL FIELD}
\noindent
In the simplest case a neutron star can be considered a solid conducting ball of typical radius $R\sim10$~km rotating with period that usually lies in the range from $P\sim1$~ms for millisecond pulsars to $P\sim10$~s for magnetars. We assume that the stellar center is fixed and the rotation axis passes through the center. The rotation axis itself is not fixed, and its direction can change in time. We do not assume that the neutron star is a perfect ball, and the shape may be arbitrary, not necessarily spherical; the stellar center in this case is defined as the zero-velocity point around which rotation of the neutron star occurs. Every point in the neutron star moves with velocity
\begin{equation}
\label{rotationVelocity}
\mathbf{v}=\mathbf{\Omega}\times\mathbf{r},
\end{equation}
where $\mathbf{r}$ is the radius vector directed from the stellar center to the point and $\mathbf{\Omega}$ is the angular velocity, with absolute value $\Omega=2\pi/P$.

Inside the neutron star we have the Maxwell equations
\begin{gather}
\label{divEin}
\operatorname{div}\mathbf{E}=4\pi\rho,
\\
\label{divBin}
\operatorname{div}\mathbf{B}=0,
\\
\label{rotEin}
\operatorname{curl}\mathbf{E}=-\frac{\partial\mathbf{B}}{\partial t},
\\
\label{rotBin}
\operatorname{curl}\mathbf{B}=4\pi\mathbf{j}+\frac{\partial\mathbf{E}}{\partial t},
\end{gather}
where $\mathbf{E}$ and $\mathbf{B}$ are the internal electric and magnetic fields, $\rho$ and $\mathbf{j}$ are the volume charge and current densities. We put $c=1$, which corresponds to measuring all velocities in units of the velocity of light, $c$, and all times in units of $1/c$, so that velocity is dimensionless and time has dimensions of length. To restore dimensions, it is sufficient to make in all formulas the substitutions $t\rightarrow c t$, $\partial/\partial t\rightarrow(1/c)\partial/\partial t$, $d/d t\rightarrow(1/c)d/d t$, $\mathbf{\Omega}\rightarrow\mathbf{\Omega}/c$, $\mathbf{v}\rightarrow\mathbf{v}/c$, $\mathbf{w}\rightarrow\mathbf{w}/c^2$, $\mathbf{E}\rightarrow\mathbf{E}/c$, $\mathbf{B}\rightarrow\mathbf{B}/c$, $\rho\rightarrow\rho/c$, $\mathbf{j}\rightarrow\mathbf{j}/c^2$, and $\sigma\rightarrow\sigma/c$.

The electric current $\mathbf{j}$ is related to $\mathbf{E}$, $\mathbf{B}$, and $\mathbf{v}$ via the relativistic Ohm law (Jackson 1999, p. 572; Goedbloed et~al. 2010, p. 590),
\begin{equation}
\label{relativisticOhmLaw}
\mathbf{j}=\sigma\gamma(\mathbf{E}+\mathbf{v}\times\mathbf{B}-\mathbf{v}\mathbf{v}\cdot\mathbf{E})+\rho\mathbf{v},
\end{equation}
where $\sigma$ is the conductivity of the neutron star and $\gamma=(1-v^2)^{-1/2}$ is the Lorentz factor and where we denote $\mathbf{a}\mathbf{b}\cdot\mathbf{c}=\mathbf{a}(\mathbf{b}\cdot\mathbf{c})$. Equation \eqref{relativisticOhmLaw} means that in the comoving inertial reference frame, which momentarily moves with the velocity of a given point, the Ohm law has the usual form
\begin{equation*}
\mathbf{j}'=\sigma{\mathbf{E}}',
\end{equation*}
where
\begin{equation*}
\mathbf{j}'=\mathbf{j}-\gamma\rho\mathbf{v}+\frac{\gamma^2}{\gamma+1}\mathbf{v}\mathbf{v}\cdot\mathbf{j}
\end{equation*}
and
\begin{equation*}
\mathbf{E}'=\gamma(\mathbf{E}+\mathbf{v}\times\mathbf{B})-\frac{\gamma^2}{\gamma+1}\mathbf{v}\mathbf{v}\cdot\mathbf{E}
\end{equation*}
are the current and electric field in this frame, by the Lorentz transform (Jackson 1999, p. 558).

We assume that the neutron star is a perfect conductor, which means that it formally has infinite conductivity,
\begin{equation*}
\sigma=\infty.
\end{equation*}
With the finiteness of $\mathbf{j}-\rho\mathbf{v}$ in Eq.~\eqref{relativisticOhmLaw}, this condition implies
\begin{equation}
\label{preForceFree}
\mathbf{E}+\mathbf{v}\times\mathbf{B}-\mathbf{v}\mathbf{v}\cdot\mathbf{E}=0,
\end{equation}
which in turn gives
\begin{equation}
\label{forceFree}
\mathbf{E}=-\mathbf{v}\times\mathbf{B}.
\end{equation}

To verify Eq.~\eqref{forceFree}, note that if it holds, then Eq.~\eqref{preForceFree} also holds because $\mathbf{E}$ is orthogonal to $\mathbf{v}$ by Eq.~\eqref{forceFree}; in other words, Eq.~\eqref{forceFree} is a solution to Eq.~\eqref{preForceFree}. We should only verify that Eq.~\eqref{forceFree} is a unique solution to Eq.~\eqref{preForceFree}. To do so, we write the electric field in the form $\mathbf{E}=-\mathbf{v}\times\mathbf{B}+\boldsymbol{\delta}$, where $\boldsymbol{\delta}$ is a deviation, and substitute it in Eq.~\eqref{preForceFree} to give $\boldsymbol{\delta}=\mathbf{v}\mathbf{v}\cdot\boldsymbol{\delta}$. We see that $\boldsymbol{\delta}$ is parallel to $\mathbf{v}$, which means that $|\mathbf{v}\cdot\boldsymbol{\delta}|=v\delta$; therefore, $\delta=v^2\delta$. Since every point in the neutron star can move only with velocity less than the velocity of light, $v<1$, we obtain $\delta=0$ and, hence, $\boldsymbol{\delta}=0$. Equation \eqref{forceFree} can also be obtained from the vanishing of the electric field in the comoving frame (Landau and Lifshitz 1988, p. 91) and coincides with the usual condition of infinite conductivity in nonrelativistic magnetohydrodynamics, when $v\ll1$ (Landau and Lifshitz 1982, p. 315); meanwhile, it is fully relativistic and is valid for every $v<1$.

Since $\mathbf{v}$ is known and given by Eq.~\eqref{rotationVelocity}, Eq.~\eqref{forceFree} determines the internal electric field via the internal magnetic field; therefore, it remains to find the latter. Substituting Eq.~\eqref{forceFree} in the Faraday law \eqref{rotEin} yields
\begin{equation}
\label{freezingInEq}
\frac{\partial\mathbf{B}}{\partial t}=\operatorname{curl}(\mathbf{v}\times\mathbf{B}),
\end{equation}
which determines the magnetic field. Equation \eqref{freezingInEq} means that every point of a magnetic field line moves with velocity of the medium at this point (Alfv\'{e}n and F\"{a}lthammar 1963); in other words, magnetic field lines are ``frozen in'' in the medium.

Taking account of Eqs.~\eqref{rotationVelocity} and \eqref{divBin} and using the known formulas of vector analysis, we obtain from Eq.~\eqref{freezingInEq}
\begin{equation}
\label{rotationBEq}
\frac{d\mathbf{B}}{dt}=\mathbf{\Omega}\times\mathbf{B},
\end{equation}
where
\begin{equation}
\label{fullDerivative}
\frac{d}{dt}=\frac{\partial}{\partial t}+\mathbf{v}\cdot\nabla
\end{equation}
denotes full, or substantial, derivative. Equation \eqref{rotationBEq} is analogous to the equation for the radius vector,
\begin{equation*}
\frac{d\mathbf{r}}{dt}=\mathbf{\Omega}\times\mathbf{r},
\end{equation*}
which reflects the equality $\mathbf{v}=d\mathbf{r}/dt$ and describes rotation of radius vector $\mathbf{r}$ around axis $\mathbf{e}_z=\mathbf{\Omega}/\Omega$ with angular velocity~$\Omega$. Therefore, the magnetic field vector at an arbitrary point in the neutron star, when parallel transferred so that its origin is at the stellar center, rotates analogously to the radius vector. Equation \eqref{rotationBEq} expresses again the freezing-in condition and describes corotation of the internal magnetic field with the neutron star.

We see that the internal magnetic field is determined at any time $t$ if it is specified at an arbitrary (say, zero) time and if the position of the neutron star is known. The latter position can be obtained from the initial position at zero time by rotation around an axis $\boldsymbol{\zeta}$, with $|\boldsymbol{\zeta}|=1$, through an angle $\alpha$, by the Euler theorem. The position can be specified by a quaternion (Branets and Shmyglevskii 1973; Zhuravlev 2001; Girard 2007)
\begin{equation}
\label{quaternionLambda}
\Lambda=\cos\frac{\alpha}{2}+\boldsymbol{\zeta}\sin\frac{\alpha}{2},
\end{equation}
which corresponds to the above rotation. Generally, $\Lambda=\Lambda(t)$, with $\boldsymbol{\zeta}=\boldsymbol{\zeta}(t)$ and $\alpha=\alpha(t)$, and describes how the position of the neutron star changes with time. We do not constrain anyhow this time dependence and consider it arbitrary.

Let us recall some necessary facts from the theory of quaternions. A quaternion $\mathrm{M}=\mu_0+\boldsymbol{\mu}$ is the sum of a scalar part $\mu_0$ and a vector part $\boldsymbol{\mu}$. The conjugate quaternion is $\bar{\mathrm{M}}=\mu_0-\boldsymbol{\mu}$; for the quaternion \eqref{quaternionLambda} we have $\bar{\Lambda}=\cos(\alpha/2)-\boldsymbol{\zeta}\sin(\alpha/2)$. Any vector is a quaternion with zero scalar part; e.g., the radius vector $\mathbf{r}$ can be considered as the quaternion $0+\mathbf{r}$. The product of two quaternions $\mathrm{M}$ and $\mathrm{N}=\nu_0+\boldsymbol{\nu}$ is defined as
\begin{equation}
\label{quaternionProduct}
\mathrm{M}\circ \mathrm{N}=\mu_0\nu_0-\boldsymbol{\mu}\cdot\boldsymbol{\nu}+\mu_0\boldsymbol{\nu}+\nu_0\boldsymbol{\mu}+\boldsymbol{\mu}\times\boldsymbol{\nu}.
\end{equation}
The product is associative, $(\mathrm{\Lambda}\circ\mathrm{M})\circ\mathrm{N}=\mathrm{\Lambda}\circ(\mathrm{M}\circ\mathrm{N})=\mathrm{\Lambda}\circ\mathrm{M}\circ\mathrm{N}$, but not commutative, $\mathrm{M}\circ\mathrm{N}\neq\mathrm{N}\circ\mathrm{M}$. The latter property reflects that two successive rotations of a rigid body generally give another position than the same rotations applied in inverse order. The radius vector $\mathbf{r}=\mathbf{r}(t)$ is related to the initial radius vector $\mathbf{r}_0=\mathbf{r}(0)$ via
\begin{equation}
\label{quaternionRotation}
\mathbf{r}=\Lambda\circ\mathbf{r}_0\circ\bar{\Lambda}.
\end{equation}

In order to find the internal magnetic field at time $t$ at position $\mathbf{r}$, we should take the magnetic field at zero time at initial position $\mathbf{r}_0$ and rotate it analogously to the radius vector; meanwhile, the initial position itself is obtained by the inverse rotation of~$\mathbf{r}$, which is characterized by the conjugate quaternion $\bar{\Lambda}$ [cf. Eq.~\eqref{quaternionRotation}]: $\mathbf{r}_0=\bar{\Lambda}\circ\mathbf{r}\circ\Lambda$. We then have
\begin{equation}
\label{BViaQuaternions}
\mathbf{B}(\mathbf{r},t)=\Lambda\circ\mathbf{B}(\bar{\Lambda}\circ\mathbf{r}\circ\Lambda,0)\circ\bar{\Lambda}.
\end{equation}

All above formulas allow $\mathbf{\Omega}$ to change its direction and absolute value in time:
\begin{equation}
\label{OmegaViaLambda}
\mathbf{\Omega}=\mathbf{\Omega}(t)=2\dot{\Lambda}\circ\bar{\Lambda}.
\end{equation}
The fact that the magnetic field \eqref{BViaQuaternions} satisfies Eq.~\eqref{rotationBEq}, which in turn implies satisfying the Maxwell equations \eqref{divBin} and \eqref{rotEin} and relativistic Ohm law for infinite conductivity, Eq.~\eqref{forceFree}, can also be verified directly, by time differentiation with account of Eq.~\eqref{OmegaViaLambda}. Thus, the internal magnetic field distribution rotates as a rigid body. This means that any corrections to the magnetic field at rest, which could appear because of rotation and be powers of~$v$, are in fact absent. Note immediately that from this follows independence of the magnetization current from the quantity $\rho\mathbf{v}$, which in turn gives breakdown of the Goldreich-Julian relation for the charge density in the neutron star (see below).

If $\mathbf{\Omega}$ is constant, then $\boldsymbol{\zeta}=\mathbf{e}_z$ and $\alpha=\Omega t$; this means that the neutron star rotates around fixed axis $\mathbf{e}_z$ with constant angular velocity $\Omega$. Equations \eqref{quaternionLambda}, \eqref{quaternionProduct}, and \eqref{BViaQuaternions} give in this case
\begin{align}
\mathbf{B}(\mathbf{r},t)=&\mathbf{B}_0(\mathbf{r}_0)\cos\Omega t + \mathbf{e}_z\times\mathbf{B}_0(\mathbf{r}_0)\sin\Omega t
\notag\\
&+ \mathbf{e}_z\mathbf{e}_z\cdot\mathbf{B}_0(\mathbf{r}_0)(1-\cos\Omega t),
\label{BForConstantOmega}
\end{align}
where $\mathbf{B}_0(\mathbf{r})=\mathbf{B}(\mathbf{r},0)$ is the internal magnetic field at zero time and
\begin{equation*}
\mathbf{r}_0=\mathbf{r}\cos\Omega t - \mathbf{e}_z\times\mathbf{r}\sin\Omega t
+ \mathbf{e}_z\mathbf{e}_z\cdot\mathbf{r}(1-\cos\Omega t).
\end{equation*}

The internal electric field $\mathbf{E}$ is given by Eq.~\eqref{forceFree}, in which we can determine $\mathbf{B}$ via Eq.~\eqref{BViaQuaternions} generally or via Eq.~\eqref{BForConstantOmega} for constant~$\mathbf{\Omega}$. In the latter case we have for $\mathbf{E}$ an equation analogous to Eq.~\eqref{rotationBEq} for $\mathbf{B}$, $d\mathbf{E}/dt=\mathbf{\Omega}\times\mathbf{E}$. Therefore, the internal electric field corotates with the neutron star when the angular velocity is constant. Equation \eqref{rotationBEq} is, however, more general and means that the internal magnetic field corotates with the neutron star even when the angular velocity changes with time. When $\mathbf{\Omega}$ is not constant, we have
\begin{equation}
\label{rotationEForArbitraryOmega}
\frac{d\mathbf{E}}{dt}=\mathbf{\Omega}\times\mathbf{E}-\mathbf{w}\times\mathbf{B},
\end{equation}
where $\mathbf{w}=\dot{\mathbf{\Omega}}\times\mathbf{r}$ is the rotational acceleration. The rightmost term in Eq.~\eqref{rotationEForArbitraryOmega} describes the change in the electric field due to the change in the velocity of the neutron star matter, which also determines the field, see Eq.~\eqref{forceFree}.

\section*{CHARGES AND CURRENTS}
\noindent
Besides the relativistic Ohm law for infinite conductivity, Eq.~\eqref{forceFree}, we have used thus far only two of four Maxwell equations to calculate and realize the behavior of the internal electromagnetic field, Eqs.~\eqref{divBin} and~\eqref{rotEin}. The remaining two Eqs. \eqref{divEin} and \eqref{rotBin} allow us to determine the underlying electric charges and currents that generate the field:
\begin{gather}
\label{rho}
\rho=\frac{\operatorname{div}\mathbf{E}}{4\pi},
\\
\label{j}
\mathbf{j}=\frac{1}{4\pi}\biggl(\operatorname{curl}\mathbf{B}-\frac{\partial\mathbf{E}}{\partial t}\biggr).
\end{gather}

Let us consider for the moment a magnetized neutron star at rest, when $\mathbf{\Omega}=0$ and $\dot{\mathbf{\Omega}}=0$. The neutron star has a distribution of the internal magnetic field $\mathbf{B}$, and the corresponding current is determined by the stationary version of Eq.~\eqref{j}:
\begin{equation}
\label{magnetizationCurrent}
\mathbf{j}_\text{m}=\frac{\operatorname{curl}\mathbf{B}}{4\pi}.
\end{equation}
Though the neutron star is at rest, $\mathbf{j}_\text{m}$ can generally be nonzero and represents a magnetization current that generates the field~$\mathbf{B}$. This is seen from the fact that at rest there are no currents related to the electric field, as then the electric field in the neutron star is zero and does not change with time.

The magnetization current is essentially linked to the internal magnetic field and exists not only at rest but also when the neutron star rotates. Since the field is frozen in, the magnetization current is also frozen in. Therefore, we have for $\mathbf{j}_\text{m}$ the same corotation equation as for $\mathbf{B}$,
\begin{equation}
\label{rotationJm}
\frac{d\mathbf{j}_\text{m}}{dt}=\mathbf{\Omega}\times\mathbf{j}_\text{m},
\end{equation}
which is valid for arbitrary rotational motion of the neutron star, when the absolute value and direction of $\mathbf{\Omega}$ can change with time. Equation \eqref{rotationJm} can also be verified directly---using Eqs.~\eqref{rotationVelocity}, \eqref{rotationBEq}, \eqref{fullDerivative}, and \eqref{magnetizationCurrent}.

Now we may consider the relation between the charge and current densities in the neutron star. Substituting Eq.~\eqref{forceFree} in Eq.~\eqref{rho} yields the charge density as a function of current,
\begin{equation}
\label{rhoJRelation}
\rho=\rho_{\text{GJ}\,0}+\mathbf{j}_\text{m}\cdot\mathbf{v},
\end{equation}
where we have denoted by
\begin{equation*}
\rho_{\text{GJ}\,0}=-\frac{\mathbf{\Omega}\cdot\mathbf{B}}{2\pi}
\end{equation*}
the so-called Goldreich-Julian density for zero velocity, into which the Goldreich-Julian density transforms when $v=0$ (at the center of the neutron star). Recall in this regard the general expression for the Goldreich-Julian density (Goldreich and Julian 1969):
\begin{equation}
\label{rhoGJ}
\rho_\text{GJ}=\frac{\rho_{\text{GJ}\,0}}{1-v^2}.
\end{equation}
where $v=\Omega r\sin\theta$ is the velocity at a given point and $\theta$ is the polar angle.

Importantly, only the magnetization current $\mathbf{j}_\text{m}$, not the total current $\mathbf{j}$, appears in the expression \eqref{rhoJRelation} for the charge density. More precisely, only the azimuthal, toroidal component $\mathbf{j}_{\text{m}\,\phi}=\mathbf{v}\mathbf{v}\cdot\mathbf{j}_\text{m}/v^2$, which is parallel to~$\mathbf{v}$, is important for the determination of the charge density. In itself, $\mathbf{j}_\text{m}$, and hence $\mathbf{j}_{\text{m}\,\phi}$, is related to the magnetic field and, evidently, is not related to the quantity~$\rho\mathbf{v}$; thus, $\mathbf{j}_{\text{m}\,\phi}\neq\rho\mathbf{v}$. Now notice that were $\mathbf{j}_{\text{m}\,\phi}=\rho\mathbf{v}$, we would obtain from Eq.~\eqref{rhoJRelation} exactly the Goldreich-Julian density~\eqref{rhoGJ} as the charge density. We immediately conclude that the charge density is not equal to the Goldreich-Julian density,
\begin{equation*}
\rho\neq\rho_\text{GJ}.
\end{equation*}

\section*{DISCUSSION}
\noindent
The charge density is thus determined not only by the angular velocity and magnetic field, as is the case of the Goldreich-Julian relation~\eqref{rhoGJ}, but also by the curl of the magnetic field. This indicates the importance of the magnetic field topology: a tube of straight magnetic field lines and a tube of twisted magnetic field lines, even with the same magnetic field strength, have different charges in the moving medium, as twisting gives nonzero curl. In other words, twisting magnetic field lines results in locally accumulating charge, and vice versa.

The charge accumulated due to twisting can play the principal role and significantly exceed the standard Goldreich-Julian value. For instance, for a closed tube of twisted magnetic field lines---a twisted torus---that lies in the equatorial plane and embraces the neutron star (see the figure), obvious order-of-magnitude estimates in Eq.~\eqref{rhoJRelation} give that the charge density is $R_0/r_0\gg1$ times $\rho_{\text{GJ}\,0}$, where $R_0$ and $r_0$ are the large and small radii for the torus. For $r_0\approx1$~km and $R_0\approx10$~km the charge density is an order of magnitude higher than the Goldreich-Julian density. The twisted torus gives an example of nonzero toroidal magnetization current not equal to~$\rho\mathbf{v}$. Note that the twisted torus is a possible structure of the internal magnetic field for the magnetar (Braithwaite and Spruit 2004; Braithwaite and Nordlund 2006; Braithwaite 2009; Mastrano et~al. 2011, 2015; Ciolfi et~al. 2011; Ciolfi and Rezzolla 2013; Lander 2013, 2014).

Twisting or untwisting magnetic field lines causes a jump or drop in the charge density and results in the appearance of currents that can heat the neutron star crust. For example, charge accumulation in the twisted torus results in the appearance of an extra electric field energy $\varepsilon\sim q^2/2R$, where $q\sim(R_0/r_0)\rho_{\text{GJ}\,0}V$ and $V\sim2\pi^2r_0^2R_0$ are the charge and volume of the torus. For $B\sim10^{15}$~G, $P\sim1$~s, and the above $r_0$ and $R_0$ we get $\varepsilon\sim10^{39}$~erg, and during the appearance or disappearance of the twisted torus we may expect the release of thermal energy of the same order. This sufficiently high energy release can be provided by deceleration of the flow of charged particles in the crust that are accelerated by a nonstationary electric field in the magnetosphere during the rearrangement, and by absorption of photons emitted by them.

Note that heating can also exist after rearrangement of the field due to formation of large stationary magnetization currents. The energy release in the crust for providing thermal emission of the magnetar is $H\sim10^{20}\text{ erg}\,\text{cm}^{-3}\,\text{s}^{-1}$ (Kaminker et~al. 2012), which for the conductivity $\sigma\sim10^{22}\text{ s}^{-1}$ gives the current density $j\sim\sqrt{\sigma H}\sim10^{21}$ cgs units. This current density can be obtained due to the electromagnetic field rearrangement accompanied by the appearance of a large charge density $\rho=\lambda\rho_\text{GJ}$, where $\lambda\gg1$. From the estimates $\rho\sim j_m v$, $j\sim j_m$, and $B\sim10^{15}$~G we obtain $\lambda\sim100$. Such a large charge density is related to a large magnetization current and indicates a change in the crustal magnetic field at small spatial scales $R/\lambda\sim100$~m. In other words, for these parameters we may assume a transition from a large-scale to small-scale magnetic field and consider it as a reason for an additional crust heating.

Thus, rearrangement of the internal magnetic field can manifest itself via crust heating by external magnetospheric effects related to radio emission. For instance, the heating can initiate a transition from a RRAT to pulsar state of the neutron star (Istomin and Sob'yanin 2011a), which has been observed in two curious hybrid radio sources, PSR J0941-39 and PSR B0826-34 (Burke-Spolaor and Bailes 2010; Esamdin et~al. 2012). Another instance concerns regular magnetars. An explanation of radio emission from magnetars was given in 2007 (Istomin and Sob'yanin 2007). It was shown that radio emission owes its existence to generation of an electron-positron plasma, and the generation is not suppressed and can occur efficiently in superstrong magnetic fields despite the quantum effect of photon splitting. In this case internal magnetic field rearrangement can switch on or intensify magnetar radio emission, because inverse Compton scattering, which increases due to heating, enhances plasma generation. A separate study is required to determine exact conditions under which the above mechanisms take place, but we can definitely conclude here that these could give a possibility to observe internal effects related to field-line twisting even in the absence of contribution from additional mechanic effects. This is of importance for the determination of the structure of the internal magnetic field for magnetars, the source of their observed activity.

The described effects of twisting magnetic field lines in a neutron star include the link between crust heating and plasma generation, or between X-ray and radio emission. Another reflection of the link was found earlier in pure RRATs: the theory (Istomin and Sob'yanin 2011a) and subsequent observations (Miller et~al. 2013) showed that physical processes producing radio bursts can heat the polar cap.

Field rearrangement requires that neutron star matter allow internal relative motions. The motions can clearly occur in a liquid matter. However, a solid matter does not rule out the possibility of rearrangement: the matter can fracture or melt, say, via the Pomeranchuk effect (Pomeranchuk 1950; Bruk 1975; Bruk and Kugel' 1976a, 1976b, 1977).

Note finally that the difference of the charge density from the Goldreich-Julian density is also important when considering electromagnetic processes occurring in the neutron star magnetosphere (Timokhin et~al. 2000; Zanotti et~al. 2012).

\section*{CONCLUSIONS}
\noindent
I have found the internal electromagnetic field of a perfectly conducting magnetized neutron star that rigidly rotates around a fixed point. The rotation axis and the absolute value of the angular velocity may change in an arbitrary way, and the velocity of matter at the stellar surface may approach the velocity of light. I have shown that the electric field corotates with the neutron star when the angular velocity is constant, while the magnetic field corotates always, even when the direction and absolute value of the angular velocity change with time. Magnetic corotation means the absence of any corrections to the magnetic field in the case of stellar rotation compared with the case of rest.

I have calculated the volume charge and current densities that generate the internal field and studied their interrelation. I have shown that the Goldreich-Julian relation in the neutron star is incorrect: the charge density is not equal to the Goldreich-Julian density. This circumstance reveals a link between charge distribution in the neutron star and magnetic field topology: the charge density is determined not only by the absolute value of the field but also by its degree of twisting. Rearrangement of the internal magnetic field is potentially observable: twisting and untwisting magnetic field lines causes internal currents that can heat the crust and change observational properties of neutron stars, such as RRATs, hybrids, and magnetars. The results of this paper can also be useful in the study of the anomalous electromagnetic torque on a neutron star.

\section*{ACKNOWLEDGMENTS}
\noindent
I would like to thank Yulii Mendelevich Bruk for useful discussions. The work was supported in part by the Educational and Scientific Complex of Lebedev Physical Institute.

\pagebreak

\centerline {\bf FIGURE}
\vspace{1cm}
\begin{figure}[h]
\begin{center}
\hspace{-2cm}\epsffile{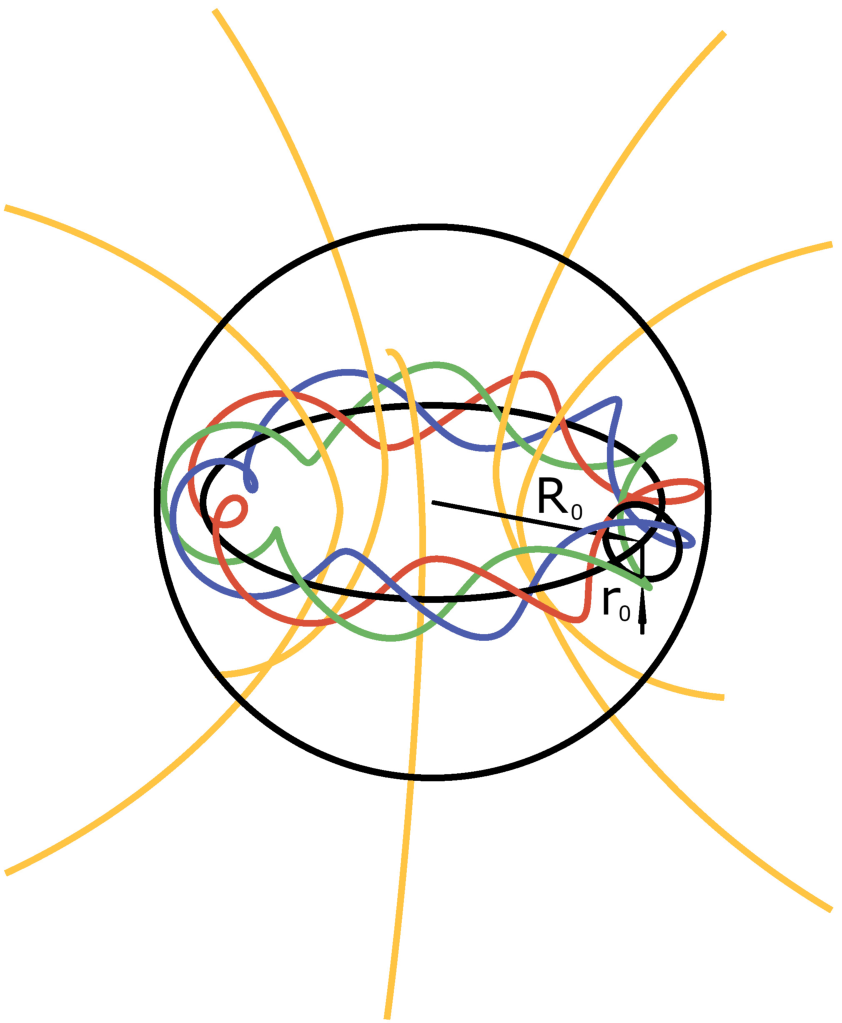}
\end{center}
\caption{Twisted torus}
\end{figure}
\end{document}